

\input phyzzx.tex
\def\ITP{\address{Institute for Theoretical Physics\break
      University of California\break Santa Barbara, CA 93106}}
\pubnum={NSF--ITP--91--105}
\date={October, 1991}
\titlepage
\title{EFFECTIVE STRING AMPLITUDES FOR HADRONIC PHYSICS}
\author{David C. Lewellen\foot{Email: DCL@SBITP.bitnet}}
\ITP
\abstract
   We propose using the general structure and properties of conformal field
theory amplitudes, in particular those defined on surfaces with boundaries, to
explore effective string theory amplitudes for some hadronic processes. Two
examples are considered to illustrate the approach. In one a natural mechanism
for chiral symmetry breaking within the string picture is proposed. One
consequence is that the vertex operator for pion emission (at zero momentum)
behaves like a world sheet current evaluated on the string boundary. This fact
is used to rederive, in a more general setting, hadronic mass relations found
in the early days of string theory by Lovelace, and Ademollo, Veneziano and
Weinberg. In the second example, we derive the general structure of the form
factor for the emission of a pomeron (interpreted as a closed string) from a
meson or baryon. The result reconciles the interpretation of the pomeron as a
closed string, emitted from the interior of the meson or baryon
world sheet, with
the additive quark rules for total hadronic cross sections. We also review the
difficulties involved in constructing complete effective string theories for
hadrons, and comment on the relation between the intercepts of trajectories
and the short distance behavior of the underlying theory.
\endpage
\REF\data{S.N.~Ganguli and D.P.~Roy\journal Phys.Rep.&67C (1980) 201;
  A.C.~Irving and R.~Worden\journal Phys.Rep&34C (1977) 117; M.~Fukugita
and K.~Igi\journal Phys.Rep.&31C (1977) 237 ;
C.B.~Chiu\journal Ann.Rev.Nucl.Sci.&22 (1972) 255.}
\REF\regge{P.D.B.~Collins, {\it Regge Theory and High Energy Physics}
(Cambridge University Press, Cambridge, 1977); P.D.B.~Collins and A.D.~Martin,
{\it Hadron Interactions} (Adam Hilger Ltd., Bristol, 1984).}
\REF\framp{P.~Frampton, {\it Dual Resonance Models} (Benjamin, 1974).}
\REF\nclim{G.~'t~Hooft\journal Nucl.Phys.&B72 (1974) 461; G.~Veneziano\journal
Nucl.Phys.&B117 (1976) 519.}
\REF\bpz{A.A.~Belavin, A.M.~Polyakov and
A.B.~Zamolodchikov\journal Nucl.Phys.&B241 (1984) 333.}
\REF\cardy{J.~Cardy\journal Nucl.Phys.
&B240 (1984) 514; {\bf B275} (1986) 200; {\bf B324} (1989) 581;
N.~Ishibashi\journal Mod.Phys.Lett.&A4 (1989) 251; J.~Cardy and
D.C.~Lewellen, {\sl Phys.Lett.} {\bf B259} (1991) 274.}
\REF\dcl{D.C.~Lewellen NSF--ITP--91--32, to appear in {\it Nucl.Phys.}
{\bf B[FS]}.}
\REF\chiral{J.~Gasser and H.~Leutwyler\journal Ann.Phys.(N.Y.)&158 (1984) 142;
S.~Weinberg\journal Physica&96A (1979) 327;
S.~Coleman, J.~Wess, B.~Zumino, and C.~Callen\journal Phys.Rev.&177 (1969)
2239; {\bf 177} (1969) 2246.}
\REF\lovelace{C.~Lovelace\journal Phys.Lett.&28B (1968) 264.}
\REF\avw{M.~Ademollo, G.~Veneziano, and S.~Weinberg\journal Phys.Rev.Lett.&22
(1969) 83.}
\REF\lenny{A.~Casher and L.~Susskind\journal Phys.Rev.&D9 (1974) 436;
R.C.~Brower and L.~Susskind\journal Phys.Rev.&D7 (1973) 1032; L.~Susskind,
A.~Casher, and J.~Kogut\journal Phys.Rev.&D8 (1973) 4448; J.~Kogut and
L.~Susskind\journal Phys.Rep.&8C (1973) 75; A.~Casher, S.H.~Noskowicz and
L.~Susskind\journal Nucl.Phys.&B32 (1971) 75.}
\REF\brower{R.C.~Brower\journal Phys.Lett.&34B (1971) 143; A.~Neveu and
C.B.~Thorn\journal Phys.Rev.Lett.&27 (1971) 1758; J.H.~Schwarz\journal
Phys.Rev.&D5 (1972) 886.}
\REF\lpom{C.~Lovelace\journal Phys.Lett.&34B (1971) 500.}
\REF\ales{V.~Alessandrini,
D.~Amati and B.~Morel\journal Nuovo Cim.&7A (1972) 797.}
\REF\cshap{L.~Clavelli and J.A.~Shapiro\journal Nucl.Phys.&B57 (1973) 490.}
\REF\qcount{D.~Sivers, S.J.~Brodsky and R.~Blankenbecler\journal Phys.Rep.&23C
(1976) 1.}
\REF\shap{J.A.~Shapiro\journal Phys.Rev.&D179 (1969) 1345.}
\REF\thick{R.~Sommer\journal Nucl.Phys.&B306 (1988) 181; A.~Di~Giacomo,
M.~Maggiore and S.~Olejnik\journal Phys.Lett.&236B (1990) 199; R.W.~Haymaker
and J.~Wosiek\journal Acta Phys.Polon.&B21 (1990) 403;
 D.G.~Caldi and T.~Sterling\journal Phys.Rev.Lett.&60 (1988)
2454; M.~Caselle, R.~Fiore and F.~Gliozzi\journal Phys.Lett.&224B (1989) 153.}
\REF\pol{J.~Polchinski and A.~Strominger, preprint UTTG-17-91.}
\REF\wip{D.C.~Lewellen, work in progress}
\REF\cremmer{E.~Cremmer and J.~Scherk\journal Nucl.Phys.&B72 (1974) 117.}
\REF\adler{S.~Adler\journal Phys.Rev.&137 (1965) B1022.}
\REF\zerot{E.~Borie et al.\journal Z. Phys.&C4 (1980) 333; R.~Odorico\journal
Phys.Lett.&33B (1970) 489; {\bf 34B} (1971) 64; {\bf 38B} (1972) 411.}
\REF\addqk{E.M.~Levin and L.L.~Frankfurt\journal JETP Lett.&2 (1965) 65;
H.J.~Lipken and F.~Scheck\journal Phys.Rev.Lett.&16 (1966) 71;
H.J.~Lipken\journal Phys.Rev.Lett.&16 (1966) 1015;\break J.J.J.~Kokkedee and
L.~van~Hove\journal Nuovo Cim.&A42 (1966) 711.}
\REF\landolt{H.~Schopper, ed.,{\it Total Cross sections for Reactions of High
Energy Particles}, Landolt--Bornstein, New Series, Vol. 12a and 12b (1987).}
\REF\donn{A.~Donnachie and P.V.~Landshoff\journal Phys.Lett.&123B (1983) 345;
{\sl Nucl.Phys.} {\bf B231} (1983) 189; {\bf B244} (1984) 322; {\bf B267}
(1986) 690.}

\chapter{Introduction}

     Originally string theory was a phenomenological theory of hadronic
physics,
developed to describe some striking features of
hadrons and their interactions.  Today, given the status of fundamental
superstrings as that branch of theoretical physics perhaps least constrained by
experiment, this history is often viewed as an ironic curiosity.
The fact remains, however, that the experimental evidence for the
string behavior of hadrons in some processes is as compelling today as it was
 20 years ago (if not more so).\foot{Relevent data can be found in several
reviews [\data]. For Regge phenomenology see [\regge], and for a summary of
some phenomenological applications of dual string models, with
references, see [\framp].}   To a reasonable approximation the
known mesons and baryons do lie on linear Regge trajectories. A wealth of high
energy, modest momentum transfer, elastic scattering data is  modelled well by
single Reggeon (open string) exchange, with the exchanged trajectories
coinciding with those found from the hadronic spectra.  Ad--hoc
phenomenological
models incorporating Regge behavior and duality (e.g., the Koba--Nielsen
amplitudes) often provide a reasonable fit with few free parameters to the
experimental data for many scattering processes.  Moreover there is
considerable theoretical prejudice (encouraged by
 the flux tube picture, and the area
confinement law found in the strong coupling expansion) that some limit of QCD
(perhaps $N_c\rightarrow\infty$) should be exactly equivalent to some string
theory.\refmark{\nclim}

      There are, however,  fundamental differences between the behavior
of QCD and that of the complete, mathematically consistent string theories
which we know how to construct.  QCD displays distinctly un--string like
behavior at short distances, where the interactions of small numbers of
point--like quarks and gluons dominate (e.g., in deep inelastic,
 or high energy fixed angle, scattering).  In all of the string models which
have been fully constructed to date,
the space--time string coordinates are represented by free massless
bosons on the string world sheet, and no qualitatively new behavior appears
even for arbitrarily short distances in space--time.  If we consider mesons as
valence quarks and anti--quarks linked by a QCD flux tube then it is clear that
the naive string picture must be modified (or break down altogether) for
processes in which the thickness of the flux tube becomes important.

      It is likely that this mistreatment of the short distance structure in
the known string models is ultimately responsible for the most persistent and
troublesome pathology of these theories from the point of view of hadronic
physics: the appearence of undesirable states in the spectrum, i.e., the
closed string trajectory with intercept two which contains the graviton and the
open string trajectory with intercept one
which includes a massless gauge boson.
We would argue that the intercepts of these trajectories are artifacts
of considering
unphysically short strings, and that the leading effect (on the long distance
physics) of altering the short distance behavior of the string
is, in fact, a shift in the intercepts of the trajectories.

      The present work is motivated by the following questions: To what
 extent can some string theory serve as an
effective theory for QCD within the kinematical regimes for which string
behavior is observed in the data and expected on general grounds?  That is, can
some consistent and tractible string theory usefully describe hadronic string
behavior if we agree to exclude processes where short distance parton--parton
interactions are known to dominate?  Or, is it the case that some corner of a
complete string model (e.g., some trajectories, the closed string sector, some
highly excited states, or loop amplitudes) cannot be neatly divorced from the
short distance physics and  will necessarily lead to undesirable
features even in the long distance predictions of the model?
In this case can we still extract sensible phenomenology
from pieces of string models
( at the level of individual amplitudes) without the existence of
 a completely
consistent theory?  Is it possible  with modern string technology to
 systematize and justify the considerable empirical sucesses of the (often
ad--hoc) Regge and dual model phenomenology of 20 years ago?

     We will not give definitive answers to these questions here. In the
next section we discuss some of the general issues and difficulties
 involved in formulating theories of hadron strings, and provide some
background and motivation for the approach employed in subsequent sections.
  In sections 3 and 4 we explore specific features of hadron
physics within the string picture but in the absence of any completely
satisfactory string model. Our chief tool is the understanding of the basic
structure and properties of conformal field theory amplitudes,\refmark{\bpz}
especially those defined on surfaces with boundaries.\refmark{\cardy, \dcl}
  The philosophy advocated is the usual one for effective theories (for example
  chiral Lagrangians\refmark{\chiral}).  We assume that many important
features of the physics are direct
consequences of the underlying symmetries involved. Any toy effective theory
with these symmetry properties built in will necessarily correctly describe
 these physical features.  We build in some physics assumptions, use our
knowledge of the general structure of string amplitudes to isolate
properties following from ``string behavior'', and postulate these as correct
physical features, notwithstanding the many problems
which exist in the toy models
considered.  Apriori there is much guess work involved in deciding what
constitutes ``string behavior''  and in deciding how certain physical
features should be incorporated into the string picture; fortunately there
exists a wealth of data from low energy hadronic physics which can be consulted
for guidance, along with the knowledge that the underlying theory is QCD.

       In section 3 we postulate a mechanism for chiral symmetry breaking
within the string picture based on the generic behavior of conformal field
theories defined on surfaces with boundaries. Some of the symmetry present in a
bulk conformal field theory is necessarily broken by the choice of boundary
conditions when world sheet boundaries are included. In this picture chiral
symmetry breaking necessarily follows from confinement and the existence of
mesons. Incorporating chiral symmetry is
in itself sufficient to guarantee the usual current algebra results for low
energy pion interactions. The incorporation of string behavior leads to
additional results. The vertex operator for pion emission behaves like a world
sheet current evaluated on the string boundary, and from this behavior follow
 mass relations between hadron states (call them
$A$ and $A^*$) which are related by S--wave pion emission,
$\alpha^\prime(M^2_{A^*}-M^2_A)=1/2$ (mod 1).\refmark{\lovelace, \avw}
These are reasonably well satisfied experimentally.

       In section 4 we consider closed strings, the pomeron,
and total hadronic cross sections. We concentrate on a feature of the data
which potentially poses a serious difficulty for the string picture.  To an
accuracy of about 5\%  total hadronic cross sections at high energy
satisfy an additive
quark rule, as if the pomeron coupled locally to the valence quarks in
each hadron.  In the string picture the pomeron is a closed string, which is
emitted from any where in the interior of the string world sheet, and there
seems to be no origin for any simple additive behavior.  In section 4 we
consider the form factor for the coupling of closed to open strings and show
how an approximate additive behavior in fact generically arises. This gives
some insight into why the naive quark model is so succesful even for some
processes where it seems inappropriate.

      As one would expect, given the attention accorded to string models of
hadrons in their first incarnation, there are few topics in this field
without direct ancestors in the early literature. Some of the background
provided in section 2 would have been considered common knowledge 15--20 years
ago, but is included here because that is no longer the case today.
  The general picture of chiral symmetry
breaking in string theory considered in section 3 can be viewed as a natural
extension of a picture originally advocated by Susskind and collaborators in
the parton string framework.\refmark{\lenny}\foot{The incorporation of current
algebra results into string amplitudes for pions was also considered from a
somewhat different point of view in [\brower].}
  The possibility that chiral symmetry incorporated
into dual models leads to mass relations between hadrons dates back to a
classic paper of Lovelace\refmark{\lovelace}, and a generalization due to
Ademollo, Veneziano, and Weinberg.\refmark{\avw}  Their discovery
within the context of Veneziano type models is
placed on a more general footing within string theory here. In addition, there
 were extensive
studies of the pomeron--meson form factor within various string models in the
early literature\refmark{\lpom, \ales, \cshap} which serve as a starting
point for the discussion of section 4.

\chapter{Effective hadronic strings}

     Consider pion--nucleon elastic scattering.  If the momentum transfer
is very large, perturbative QCD is valid, and we can view the event as the hard
scattering of a quark or anti--quark in the pion off of a single quark in the
nucleon, with additional hard gluons exchanged within the individual hadrons to
prevent their fragmentation. The fixed angle scattering amplitude falls off
as a power of the center of
mass energy, $A\sim s^{-3}$, by the dimensional counting
rules.\refmark{\qcount}  On the other hand,
when all of the momenta involved are small, pions and nucleons behave like
point
particles.  An effective Lagrangian which incorporates an approximate
(spontaneously broken) chiral symmetry gives a good description of the
scattering event.

    An effective string theory description for pion--nucleon scattering
 is useful if there
exists some intermediate energy regime in which the scattering process
is dominated by the exchange of extended, string--like, configurations. In this
picture $\pi$N scattering is dominated by the exchange of the
$\rho$--trajectory (the $\rho$\ meson together with the infinite tower of
excited ``$\rho$--strings'' with ever increasing angular momentum), and the
amplitude exhibits the corresponding Regge behavior, $A\sim
s^{\alpha_\rho(t)}$, for large $s$.
  The principle evidence for this string behavior is the
linearity of the $\rho$--trajectory function, $\alpha_\rho(t)$, as determined
from meson spectroscopy (for $t>0$) and from charge exchange $\pi$N scattering
(for $t<0$) at energies up to tens of GeV and momentum transfers of a few GeV
or
less.  It is data of this sort which an effective string theory for hadrons
should address most directly.

     Within what regime should the effective string picture be valid?  More
specifically, when does string perturbation theory begin to break down (i.e.,
loops grow in importance) and when does the finite thickness of QCD flux tubes
become significant?\foot{These issues are not unrelated in that the string
world sheet can develop some effective thickness via a large number of small
holes and handles. In an effective theory, however, we should include
the short distance effects (to the extent possible) into the tree--level
theory, and include only loop amplitudes defined on a larger distance scale.}
At least within the Regge limit (large $s$, fixed $t$) we have some indication
 of when multiple string exchange becomes important. If the contribution
 from single trajectory exchange behaves for large $s$ like $\sim
 s^{\alpha^\prime t+\alpha_0}$, then the contribution from exchanging two
such trajectories behaves like $\sim s^{\alpha^\prime
t/2+2\alpha_0-1}$.\refmark{\regge}  Thus the loop contributions become
more important as $s$ and
$|t|$ grow.  This effect is apparent in the $\pi$N scattering data. For large
$s$ ($\sim$ 200 GeV$^2$) the measured effective trajectory remains linear for
small $|t|$ but flattens out for $t$ around $-1$ GeV$^2$.

   It should be kept in mind that the string coupling constant is not obviously
small.  Comparing the current algebra result for $\pi\pi$ scattering with the
Lovelace--Shapiro--Veneziano four--point open string
amplitude\refmark{\lovelace, \shap} leads one to conclude that the string
coupling is of the order $g\approx (\sqrt{\alpha^\prime}4\pi f_\pi)^{-1}\sim
1$.
It is only by virtue of kinematic factors (and factors of 2$\pi$) that the loop
amplitudes are suppressed for some processes. This is analogous to the
situation in chiral perturbation theory,
which the string should match on to at low energies.

   The effects of finite string thickness are more difficult to address.
The fundamental dimensionful parameter in the theory is the string tension
 $T$, or the related Regge slope, $\alpha^\prime=(2\pi T)^{-1}$.
The latter, as measured from the rho trajectory, is .88 (GeV)$^{-2}$,
corresponding to a string tension of .91 GeV/fm.  In point particle
theories large momenta are naturally associated with small distance
scales. In a string theory, on the other hand, large momenta are naturally
 associated with
{\it large} distances, via the string tension. Put simply, large momenta are
 typically exchanged via long strings, small momenta by small ones.
This basic fact leads to
a fundamental difficulty: in momentum space there is no simple way  to
identify or isolate
processes involving  short distance scales where the string picture breaks
down. This scale, essentially the flux tube thickness,
we expect (from lattice simulations and the size of heavy q$\bar{{\rm q}}$
bound states\refmark{\thick})  to be of the order of
.3 fm. It is inappropriate to eliminate such scales with a momentum cutoff
of p$\ll$1/.3 fm $\approx$ .7 GeV as one would do in an effective point
particle theory.  Not only would this
eliminate precisely  the regime which we are most interested in (where
stringlike behavior is observed in the data), but it would not insure that
short distance scales have been removed,
since small momenta can signal the appearance of unphysically short strings
(shorter than .3 fm).

     Directly imposing a short distance cutoff in space--time is no less
problematic. The utility of an effective theory depends on its calculability;
with current technology this restricts us to  string theories
constructed out of conformal field theories. By virtue of the conformal
invariance, however, there is no direct relation between a short distance
cutoff on the world sheet and one in space--time. Choosing a parametrization
of the world sheet which behaves otherwise and then imposing a cutoff
necessarily breaks the conformal
invariance, rendering the sum over string world sheets virtually intractable.

    These short distance problems are fundamentally connected with the most
serious of the traditional problems plaguing string theories applied to
hadronic physics.  The consistent theories known 15 years ago, the bosonic and
Ramond--Neveu--Schwarz (RNS) strings, suffer from a number of problems.  The
intercepts of the leading trajectories in the closed and open string sectors of
the theory are 2 and 1, respectively, instead of $\sim$1 and $\sim$1/2 as
required for the observed pomeron and $\rho$--$\omega$--A$_2$--f trajectories;
the simple bosonic (respectively RNS) string is consistent only in 26 (10)
space--time dimensions; finally, the string amplitudes do not exhibit any
parton--like behavior at short distances (e.g., power law fall off of the
elastic scattering amplitude at high energy and fixed angle).

   The last of these difficulties we have chosen to set aside in considering
the string only as an effective theory.  The first, the intercept problem, is
critically important for the physics we wish to study. The ``critical
dimension'' problem is, in comparison, only of secondary importance, and the
name is a misnomer.  What is constrained is not the number of space--time
dimensions, but a measure of the number of degrees of freedom living on the
string world sheet, the total central charge. It is easy to construct
consistent strings in four dimensions provided we include internal degrees of
freedom in the form of a conformal field theory with the appropriate central
charge; the issue is then which sort of conformal field theory is most
appropriate for hadronic applications (e.g., bosonic, fermionic,
Liouville--like as in so--called ``non--critical'' strings, etc.). While the
intercept problem appears at tree--level, a complete specification of the
degrees of freedom in the string can be postponed until loop--level.
 This will be our philosophy here, motivated by the observed string
behavior in hadronic interactions. Because the u and d quark masses are small,
hadron strings are easily broken, and the observed string behavior is dominated
by ``short'' strings like the lowest few resonances on the $\rho$ or K$^*$
trajectories.  On the other hand, in a world with all quark masses
$\gg\Lambda_{{\rm QCD}}$, we could concentrate on long strings
(length$\gg$.3 fm)
for which an intercept shift of 1 or 1/2 is completely negligible.  In this
case (which is the one discussed in [\pol]) the intercept problem would become
secondary and the specification of the string degrees of freedom would take
precidence.

    The origin of the intercept problem is the following. In the known
consistent string theories the space--time coordinates are realized in terms of
free bosons on the string world sheet.  All open string vertex operators
carrying momentum $k$ are of the form,
$$V_\phi(\zeta,k,x)=\zeta^{\mu_1\dots\mu_J}\partial X_{\mu_1}\dots
\partial X_{\mu_J}e^{ik\cdot X(x)}\phi(x)\quad ,\eqn\vphi$$
where $\phi(x)$ is an operator in the internal conformal field theory. Like
$\psi_\mu$ in the RNS model, $\phi$ need not be a Lorentz scalar, but for
notational simplicity we assume that is the case in the following equations.
The mass shell condition is that $V$ have conformal dimension one,
$J+\Delta_\phi-\alpha^\prime M^2=1$.  The intercept of the leading trajectory
of particles of this type is then,
$${\rm open\ strings:\quad }\alpha_0=1-\Delta_\phi\quad . \eqn\ointc$$
Closed string vertex operators have the form
$V(k,z,\bar{z})=V(k/2,z)\bar{V}(k/2,\bar{z})$ and have conformal dimension
(1,1) on shell so that,
$${\rm closed\ strings:\quad }\alpha_0=2-\Delta_\phi-\bar{\Delta}_{\bar{\phi}}
        \quad . \eqn\cint$$

   It is simple enough to choose internal conformal field theories with
desirable operators $\phi$ to obtain trajectories with practically any
intercept we might wish. The point is that regardless of which conformal field
theory is chosen, it will always contain the identity operator, which has
dimension (0,0).  Hence there will always be leading open and closed string
trajectories with intercepts 1 and 2. In string theories with world sheet
supersymmetry, the lowest states on these trajectories (both tachyons) in fact
decouple from physical states, but finding a symmetry such that an entire
{\it trajectory} decouples is an entirely more formidable task.

   How is this problem related to that of cutting off or modifying the string
at short distances? The following heuristic argument gives some idea of how the
parameters in an effective theory are altered when the short distance behavior
(in space--time) is modified. To implement the modification,
let us allow the string tension to change
with momentum scale, but constrained so that $T(p)$ is analytic in
 $p$, Lorentz invariant, and asymptotically constant for large $p^2$,
$${1\over 2\pi T(p)}\approx \alpha^\prime +{a\over p^2}+{b\over (p^2)^2}+\dots
\eqn\alp$$
As noted previously, it is appropriate for most string applications to hadron
physics to associate large $p^2$ with large distances and hence with
asymptotically constant string tension. This expression is valid only for
sufficiently large $p^2$ (long distances) where the string picture is still
appropriate. Given \alp\ the conformal dimension of
that piece of the vertex operator carrying momentum is altered, and with it the
behavior of the leading Regge trajectory,
$$\alpha_{{\rm leading}}(-p^2)=1-\Delta(e^{ip\cdot X})=
1-{p^2\over 2\pi T(p)}=1-a-\alpha^\prime p^2-{b\over
p^2}+\dots\eqn\shift$$
In other words,  there is an effect of modifying the small $p^2$ (short
 distance) piece of the theory, even for arbitrarily large $p^2$ (long
distances) which
is precisely to shift the intercepts of all of the trajectories by a constant
$a$.  The higher order corrections curve the trajectories for smaller $p^2$.

   This argument links two basic problems but solves neither.  Three possible
approaches come to mind: 1) find an alternative to free bosons for
incorporating string coordinates which differs at short distances but remains
conformal on the world sheet; 2) find a tractable way to sum over world sheets
for non--conformally invariant theories; 3) restrict attention to pieces of
string models (built from conformal field theories) in which the undesirable
trajectories do not appear.

    The first possibility would be ideal, allowing a systematic treatment even
at the loop level; however, such a theory is tightly constrained and its
existence can largely be ruled out, at least for theories which share the usual
ghost structure of the bosonic or RNS strings.\refmark{\wip}
  The second possibility requires new technology and may prove to be
solvable only numerically. In the remainder of
this work we will explore the third possibility, which should be appropriate
for some processes whether or not either of the first two possibilities are
realizable. Even if the complete string theory is not conformally invariant, we
expect that for some processes the string amplitude should be dominated by the
long distance degrees of freedom so that an effective conformal theory should
be appropriate for computing that particular amplitude.

    In an effective point particle theory one typically specifies a Lagrangian
with some number of interaction terms whose form is constrained by a set of
imposed symmetries. The various coupling constants are fixed by comparing the
predictions of the theory with enough measured processes for which the
effective description is valid. This done, predictions can be made for
other processes and tested against experiment.  Not much is learned if we are
restricted to consider only a single amplitude or set of amplitudes, because
there is considerable freedom in adding interaction terms and tuning coupling
constants. An effective string theory in a sense suffers from the opposite
problem.  The symmetries imposed (world sheet conformal invariance, duality,
etc.) are so constraining that we have not been able to formulate a complete
theory in which we can tune the parameters to their physical values. To some
extent, however, we can turn this fact to our advantage: precisely because the
string symmetries are so constraining we have a chance of obtaining significant
predictions from the effective theory even at the level of individual
amplitudes.

    The principle tool in this approach is the known general structure and
properties of conformal field theory amplitudes. In the following sections of
this paper we will demonstrate this approach with two examples; we will
address only a few general points here.
 In a given amplitude we must
specify the conformal dimensions of the states appearing and their fusion
rules.\foot{The operators may be organized under some  chiral
algebra extended beyond the Virasoro algebra, but for present purposes we don't
need to know explicitly what that algebra is.} Conformal invariance, duality
and possibly other symmetries for each given case, then typically fix the form
of the amplitude completely up to some coupling constants. The complexity of
the
amplitude and number of undetermined coefficients is governed primarily by the
number of distinct primary fields which appear as intermediate states.  The
lowest order approximation to a four--point function, for example, assumes that
only a single trajectory (and its daughters) is exchanged in each ($s$, $t$ and
$u$) channel; this gives amplitudes which are finite sums of Veneziano terms
and often leaves only the overall normalization undetermined. These were the
amplitudes available for phenomenological study 20 years ago. If two distinct
trajectories appear in each channel then the basic amplitudes are integrals of
hypergeometric functions; more than two leads to integrals of generalized
hypergeometric functions, etc..

   Unless otherwise noted, we take the momentum dependence of the conformal
dimensions of vertex operators to be as in the bosonic string,
$\Delta=-\alpha^\prime p^2+\dots$.
This is required to obtain the usual Klein--Gordan propagator. This,
together with momentum conservation, fixes the momentum dependence of
the conformal field theory amplitude to be as in the bosonic string, except for
the possible momentum dependence residing in coupling constants (OPE
coefficients). The conformal field theory amplitude must still be integrated
over the appropriate moduli space of vertex operator positions to obtain the
final string amplitude, so the final momentum dependence can be much different
than found in bosonic string amplitudes.

   When the conformal field theory amplitude is defined on a surface with
boundaries we need (in addition to conformal dimensions and fusion rules) some
information on the boundary conditions.  The structure of the amplitude is
correspondingly richer than in the bulk case. The necessary technology and
notations employed here are collected in [\dcl]. It is through the boundary
conditions that the valence quark properties of mesons should be incorporated.

   Without a complete string theory we have at best only some general
guidelines for determining when an effective string amplitude should be
appropriate. Processes which are obviously dominated by short distance physics,
for example large momentum transfers via electroweak currents, or heavy
q$\bar{{\rm q}}$ bound states, should be avoided. On the other hand
the interactions of hadrons
with soft electroweak probes, where vector meson dominance is appropriate,
should be amenable to a string description, and light--heavy q$\bar{{\rm q}}$
states are a particularly interesting forum for these methods.  As a practical
matter the upper limit on momentum transfers which can be considered in soft
hadronic processes should be determined by the expected onset of string loop
corrections, rather than by the appearance of short distance physics. The
possible need for a low momentum cutoff to avoid unphysically short strings is
avoided in choosing by hand the intercepts of the exchanged trajectories.

   Finally we come to hadronic  states which are easily treatable within the
string framework for some purposes but not for others. Consider again $\pi$N
scattering. The exchanged $\rho$ trajectory is modelled  adequately by an
open string. The observed string--like behavior for this amplitude says little,
however, about the validity of interpreting the pion and nucleon as one
dimensional strings. In this process they appear just
as external point sources on the string world sheet. We expect some
difficulties with the string interpretation in both cases. The pion is the
smallest of hadrons built from
light quarks, and so one might expect short distance effects (in
particular spin--spin interactions) to alter the simple string behavior on the
$\pi$ trajectory. To the extent one can judge from only three observed states
($\pi,\ b_1$ and $\pi_2$) and modest $t$ channel exchange data, this appears to
 be the case. The slope of the trajectory for $t$ near 0 is shallower than
that of the $\rho$ trajectory, but curves upward so that between the $b_1$ and
 $\pi_2$  the two trajectories are almost parallel. To model an amplitude
involving the pion trajectory with an effective string amplitude we must decide
how best to approximate this curved trajectory. In section 2 we will
concentrate on the pion as a Goldstone boson, and accordingly treat the $\pi$
trajectory as linear with the universal Regge slope, and with its intercept
near zero. This allows for the correct $\pi$ mass, but mistreats the higher
lying states on the trajectory.

    Baryons are problematic within the string picture. Generalizing the single
flux tube picture for mesons, it is natural to picture flux tubes eminating
from each of the three quarks in a baryon and joining at a point. The baryon is
then ``Y'' shaped and its world sheet in space--time consists of three
subsheets
sewn together along a joining curve, with appropriate quark boundary conditions
on the other edges (fig.1). In principle to compute an amplitude involving
baryons we must compute the corresponding conformal field theory amplitude
defined on this world sheet and then integrate over some moduli space of
joining curves as well
as over the positions of vertex operators. In the case of a meson tree
amplitude we could (by the Riemann mapping theorem) always conformally map
 the world sheet into, for example, the unit disk or upper half--plane. For the
three sheeted baryon world sheet there is no such simple result, and so the sum
over world sheets remains formidable even after the nontrivial computation of
the relevent amplitude on this surface has been performed.

   To make computations involving baryons tractable we must reduce
baryon world sheets to the form  of meson ones.  In the examples considered in
this paper we will use two different approaches. If, in the amplitude being
considered, there are no vertex operator insertions in the interior or on the
boundary of one of the three subsheets of the baryon world sheet, then we can
imagine first summing over all of the possibilities for that subsheet.  What
remains will be the computation of a correlation function in a (probably
modified) conformal field theory on a meson--like world sheet. The second
possibility is to consider the baryon as an open string with a boundary
condition appropriate to a quark on one end, and some different  boundary
condition mocking up a ``diquark'' at the other end. This approach is
particularly suited to processes, such as meson emission, which only
directly involve one of the three baryon subsheets; the details of how the
other two subsheets are treated should be of secondary importance.  Obviously
we have no systematic handle on the validity of either of these approaches to
baryons. Both are suspect if heavy quarks are involved. The best we can do is
to
compare the results of both approaches to see when they are compatable, and to
determine how sensitively they depend on the details of the boundary conditions
or conformal field theories considered.

\chapter{Chiral symmetry breaking and the nature of the pion}

      The incorporation of chiral symmetry into a low energy effective
Lagrangian is sufficient by itself to guarantee all of the current algebra
results for soft pion amplitudes. If we assume in addition
that in some intermediate energy
regime pion amplitudes exhibit string behavior, do any other general
predictions
follow?  At this stage we cannot construct a completely satisfactory string
theory which reduces to the non--linear sigma model in the low energy limit.
We can, however, consider string models with spontaneously broken symmetries,
 explore the general features of the resulting Goldstone bosons, and hope that
we are getting some of the general features of the physics correct.

    Consider then a simple toy string model: an orientable bosonic string with
string coordinates giving rise to 4--dimensional space--time together
 with some internal
conformal field theory (with $c=22$) which includes an SU(2) WZW model as one
piece.  The remainder of the internal degrees of freedom we leave unspecified
as they will play no role in the discussion.  This model contains conserved
SU(2)$\otimes$SU(2) Kac--Moody currents,
$J^a(z)$ and $\bar{J}^a(\bar{z})$, on the string
world sheet.  These are naturally associated with chiral
SU(2)$_L\otimes$SU(2)$_R$ currents in space--time (in momentum space),
   $$\eqalign{{\it J}^a_{\mu,L}(k)&=\int d^2 z J^a(z)\bar{\partial}X_\mu
         e^{ik\cdot X/2}\cr
        {\it J}^a_{\mu,R}(k)&=\int d^2 z\bar{J}^a(\bar{z})\partial X_\mu
         e^{ik\cdot X/2}\cr}\eqn\vlvr$$

   For closed strings on shell (i.e., $k^2=0$ so that the above operators are
conformally invariant on the string world sheet) these are recognized as the
vertex operators for the emission of massless SU(2)$_L\otimes$SU(2)$_R$ gauge
bosons coupling to the conserved currents in space--time.  Ultimately,
closed string gauge bosons are undesirable for hadronic physics
(and are in part a manifestation of the omnipresent leading trajectory
problem), but for our modest purposes at present it is easy to sidestep this
problem.  At tree level we can isolate the closed string sector of the model
from the open strings which are of primary interest; in fact this model can be
adjusted to eliminate these closed string gauge bosons from the spectrum
altogether without altering the open string sector of the theory.  For the
moment, though, the presence of this local symmetry will actually prove helpful
in diagnosing the nature of the symmetry breaking in this model.

      In the closed string sector of this model both of the space--time
currents are exactly conserved,
$$\eqalign{k\cdot{\it J}^a_L&=\int d^2z J^a(z)k\cdot\bar{\partial}X
   e^{ik\cdot X/2}=\int d^2z \bar{\partial}[J^a(z)e^{ik\cdot X/2}]=0\cr
          k\cdot{\it J}^a_R&=\int d^2z \bar{J}^a(\bar{z})k\cdot\partial X
   e^{ik\cdot X/2}=\int d^2z \partial[\bar{J}^a(\bar{z})
    e^{ik\cdot X/2}]=0\cr}\eqn\kv$$
The integrals of the total derivatives identically vanish upon integration
by parts since closed string world sheets have no boundaries.

     Once we consider world sheets with boundaries, that is open string
``meson'' amplitudes, we find that at least half of this symmetry must be
broken, depending on the boundary conditions chosen.  We can always conformally
map an open string tree amplitude to the upper half plane with the real axis as
boundary.  If we choose boundary conditions $J^a(x)=\bar{J}^a(x)$ on the real
axis then $\bar{J}^a(\bar{z})$ is the analytic continuation of $J^a(z)$ into
the lower half plane.  The familiar derivation  of the Ward identities
 following from a  world sheet symmetry (in terms of contour integrals
of $J^a(z)$)can be applied as usual but only for a single SU(2)
symmetry.\refmark{\cardy}
The surviving symmetry is the diagonnal SU(2) within SU(2)$_L\otimes$SU(2)$_R$,
and all boundary operators (i.e., the vertex operators for open strings) fall
into representations of this single SU(2), not the full symmetry present in the
bulk.  More generally we could choose boundary conditions preserving some other
 SU(2) (i.e., $J^a(x)=M^{ab}\bar{J}^b(x)$ with $M^{ab}$ an SO(3) rotation
matrix), or even break the symmetry completely, but no more than half of the
original symmetry can be preserved once boundaries are included.

     This is a generic feature of conformal field theories defined on surfaces
with boundaries.  If there are local world sheet symmetries present in the
bulk theory, only a subset can be preserved in the presence of boundaries.
The canonical example is the local conformal symmetry itself; the bulk
(closed string) theory includes two independent Virasoro algebras, while the
theory with boundaries (open string) includes only one.

      What is the nature of this symmetry breaking in space--time?  The simple
string model we are considering can be constructed free from any world sheet
 sickness (i.e.,
violations of modular invariance) and so should have a consistent space--time
interpretation.  Since half of the local gauge symmetry present at closed
string tree level is broken when we couple in open strings, the corresponding
gauge bosons will acquire a mass.  The only candidates for the extra degrees of
freedom required to make these massive are the massless {\it open}
 string states with vertex operators,
$$V^a_G(k)=\int dx J^a(x)e^{ik\cdot X(x)}\quad\quad .\eqn\vg$$

     This is the direct analogue (for gauge bosons in lower dimensional open
string models) of the Cremmer--Scherk version of the Higgs
mechanism.\refmark{\cremmer}
The lost symmetry is effectively spontaneously broken, with the open string
states \vg\ playing the role of the would--be Goldstone bosons.  These states
mix with one combination of the closed string vector boson vertex operators,
$${\it A}^a_\mu(k)\equiv\int d^2 z[J^a(z)\bar{\partial} X_\mu
-\bar{J}^a(\bar{z})\partial X_\mu]e^{ik\cdot X/2}\quad\quad .\eqn\va$$
The other combination of vector bosons remains massless,
$${\it V}^a_\mu(k)\equiv\int d^2 z[J^a(z)\bar{\partial} X_\mu
+\bar{J}^a(\bar{z})\partial X_\mu]e^{ik\cdot X/2}\quad\quad .\eqn\vv$$
 If we include the interchange of left and right moving degrees of freedom on
the world sheet (i.e., holomorphic and anti--holomorphic), in defining the
action of the space--time parity operation on the world sheet fields, then
${\it A}^a_\mu(k)$ and ${\it V}^a_\mu(k)$  transform as axial--vector and
vector currents respectively.

    If we just compute at open string tree level and don't consider
 the closed string sectors of this model, then the states in \vg\ forestall
being eaten and behave exactly like true Goldstone bosons arising from a
spontaneously broken SU(2) chiral symmetry.  In particular the current
algebra/soft pion theorems will be satisfied.  From \vg\ and \va\ the coupling
of the (off shell) axial vector current to the pion can be computed,
$$\langle 0 \vert g{\it A}_\mu^a(k)\vert \pi^b(p)\rangle
  =ig\delta^4 (p-k)\delta^{ab}p_\mu\quad\quad .\eqn\vavg$$
g is the string coupling constant, which we see is inversely proportional to
$f_\pi$ (as we would have found as well by comparing the amplitude for
$\pi\pi$ scattering with the current algebra result). Contracting \vavg\ with
$p^\mu$  we would find the usual PCAC relation between the pion field and the
divergence of the axial current were our pion not
exactly massless.

    The basic soft pion theorem (the Adler consistency
condition\refmark{\adler}) which states that amplitudes involving
pions should vanish as the pion momentum is set to zero and any pole terms are
removed, follows simply as well.  The vertex operator \vg\ which inserts a pion
 can be written  at zero momentum as $\oint J^a(z)dz$ with the contour
running around the world sheet boundary.  Any tree level open string amplitude
 including this vertex
operator will be an analytic function of $z$ everywhere inside of this contour
(and hence will vanish upon integration) except for possible poles when $z$
sits
atop one of the other vertex operators on the boundary. But the contributions
from these points are precisely pole terms in the space--time amplitude.

     The fact that world sheet axial currents which are conserved in the
interior of the world sheet but not on its boundaries imply the usual
current algebra results was discussed in considerable detail by Susskind and
collaborators within the parton string picture long ago.\refmark{\lenny}
The explicit toy model
considered here adds one new feature to the general scenario considered
there: the vector and axial vector currents are intimately related on the
world sheet, being different combinations of the same holomorphic and
anti--holomorphic currents; as a consequence, the presence of a boundary
necessarily breaks half of the symmetry (this needn't be put in by hand).  In
other words, in this picture confinement (assumed {\it ab initio} in the string
picture) together with flux tubes with ends (mesons) necessarily imply the
breaking of chiral symmetry.

     In its details our toy model is inadequate to describe physical
pions (there is, for example, a massless, spin one, isoscaler which couples to
$\pi\pi$); nonetheless, we expect some features to survive in any realistic
hadronic string model.  In particular we expect the vertex operators for zero
momentum pions
to behave like currents evaluated on the world sheet boundary.  In conformal
field theory language this implies (among other things) that the zero momentum
pion operator has simple fusion rules with other operators in the theory and as
a consequence its correlation functions have a simple analytic structure.  We
will show momentarily that this leads to mass relations between hadronic states
linked by S--wave pion emission, but first we must consider what sort of
modifications of our toy model we should expect and allow for.

     So far we have considered isospin and chiral--isospin  currents living in
the interior of the string world sheet.  As we have seen, the chiral
symmetry breaking is particularly clear (and automatic) in this case; however,
the existence of these currents is problematic. Most glaring is the
 presence of gauge
boson vertex operators in the closed string sector which  promote
 the global symmetry into a local one in space--time. In addition we have
closed
string states carrying nonzero isospin and exotic states with isospins greater
than one. Finally, the intuition from the large N limit of QCD (which we expect
to be string--like) is that world sheets with flavor currents in the interior
should be suppressed, since flavor separations in the interior of the
world sheet imply color separations as well.

     To what extent should flavor degrees of freedom be represented in the
interior of the string world sheet, or restricted to its boundaries?  It is
reasonable to expect that at long enough distances an effective string picture
should be valid for a gauge theory with any values of $N_c$, $N_f$, and quark
masses consistent with confinement.  This needn't be the same string theory for
different values of these parameters, and presumably some limits lead to much
simpler theories than others.

   Very heavy quarks ($M^2\gg\Lambda^2_{QCD}$) can
only appear via (non--trivial) boundary conditions on the edge of
the world sheet.  Light quarks,on the other hand,
 at least renormalize the string tension. For
$N_c$ very large this is probably their only effect on the interior of the
world sheet, and the individual flavor degrees of freedom should be present
only on the boundaries (e.g., via Chan--Paton factors).  At the other extreme,
if  $N_c$ is small,
quark flavor currents in the interior are probably not supressed
(especially if $N_f$ is relatively large, but consistent with confinement)
.  In between
these limits there should exist conserved flavor currents, which effectively
have some limited extent of penetration into the interior of the world sheet.

   While these different limits represent quite different physics in the bulk
of the world sheet, we expect much of the boundary physics to be the same.
That is, regardless of how free they are to penetrate in the interior, near the
world sheet boundary we expect conserved isospin and chiral--isospin currents,
with the latter broken as before by the boundary conditions. The pion vertex
operator should again be a world sheet current operator evaluated at the
string boundary, even if these currents do not actually propagate freely in the
bulk.

     Since they are ultimately only evaluated on the world sheet boundary, the
nature of these currents can be somewhat more general than in our toy example.
We must, in particular, allow for half--integer spin currents as well as
integer
ones.  We need only require that correlation functions involving these currents
are well defined and single valued, and this can be achieved with fermionic
boundary currents.  There is no analogue for open strings of the modular
invariance requirement under $\tau\rightarrow\tau+1$ which forbids
 operators with half--integer spin in the bulk theory.

    As an aside, we can use the partial independence of bulk and boundary
theories to modify our
toy model to remove the closed string gauge bosons. The simplest example which
illustrates the possibilities are the allowed combinations of two Ising models
(i.e., two Majorana fermions and their associated spin operators). There are
two consistent, modular invariant theories which can be defined in the bulk
theory: the naive tensor product of two Ising models, and a correlated tensor
product which includes holomorphic and anti--holomorphic fermion bilinears
which generate a U(1)$\otimes$U(1) symmetry.  When we include boundaries into
the latter theory, at most the diagonal U(1) survives; the other combination of
fermion bilinears will mix with the ``would--be Goldstone'' boundary operator
given by the fermion bilinear on the boundary.  This same operator is
present in the simple Ising$\otimes$Ising theory, even though the bulk U(1)
currents are not. In the simple Ising$\otimes$Ising theory there also
exist operators in the boundary theory which explicitly break the U(1)
symmetry (the boundary spin operator of a single one of the Ising models).
 We can, however, exclude these operators from the theory if we so choose, by
excluding some of the possible boundary conditions. There is no analogue of
modular invariance for open strings which forces us to keep all of the allowed
 boundary
conditions and operators.\refmark{\dcl}  In this truncated system of boundary
operators we have a U(1) symmetry. We could consider SU(N)$\otimes$SU(N)
  Kac--Moody currents constructed from  free majorana fermions in the
 same fashion. Or U(1) currents can be extended to non--abelian ones on the
boundary with the addition of Chan--Paton factors.

      We are now in a position to place one of the more intriguing results from
the early days of string theory on a more general footing. In 1968
Lovelace\refmark{\lovelace}
(and independently Shapiro\refmark{\shap})
considered a simple model of the Veneziano type for $\pi^+\pi^-$ scattering,
$$A(s,t)=\beta{\Gamma(1-\alpha^\prime s -\alpha_\rho)\Gamma(1-\alpha^\prime t
 -\alpha_\rho) \over \Gamma(1-\alpha^\prime s -\alpha^\prime t-2\alpha_\rho)}
\quad\quad .\eqn\ls$$
Lovelace noticed that the Adler consistency condition (vanishing of $A$ at
$s=t=u=M^2_\pi$) is satisfied given the physical value of $\alpha_\rho$, the
intercept of the rho trajectory. Or, conversely, imposing the Adler
condition forces $\alpha^\prime(M_\rho^2-M_\pi^2)$ to be an integer or
half--integer. Subsequently, Ademollo, Veneziano and Weinberg\refmark{\avw}
(AVW) generalized the argument to amplitudes of the form $\pi A\rightarrow BC$,
assuming each amplitude to be a
sum of Veneziano terms with each term vanishing at the Adler
point. For an S--wave coupling of the pion to $A$ to form an intermediate
resonance $A^*$ ( and similarly $B$ to $B^*$, etc.)
the Adler condition is nontrivial and leads to a quantization
condition of the form $\alpha^\prime(M^2_{A^*}-M^2_A+M^2_{B^*}-M^2_B)=$\
integer. Comparing various different amplitudes and using the observed
$\pi-\rho$\ splitting ($\alpha^\prime(M^2_\rho-M^2_\pi)=1/2$) then leads to the
relation,\refmark{\avw}
$$\alpha^\prime(M^2_{A^*}-M^2_A)=1/2\quad ({\rm mod\ }1)\quad.\eqn\avwr$$

   Comparing the measured mass squared differences in units of the string
tension for pairs of hadrons
linked by the emission of an S--wave pion we find:
$$\eqalign{&{A^*-A\over \hphantom{XXX}}\quad\quad\quad
            {\alpha^\prime(M^2_{A^*}-M^2_A)\over \hphantom{XXX}}\cr
           &\rho-\pi \quad\quad\quad\quad\quad\quad.50\cr
           &K^*-K \quad\quad\quad\quad\quad.49\cr
           &D^*-D \quad\quad\quad\quad\quad.48\cr
           &B^*-B \quad\quad\quad\quad\quad.49\cr
           &\Delta-N \quad\quad\quad\;\quad\quad.56\cr
           &\Sigma(1385)-\Lambda \quad\quad\quad.59\cr
           &\Sigma(1385)-\Sigma \quad\quad\quad.44\cr
           &\Xi(1530)-\Xi \quad\quad\quad.54\cr
           &\Sigma_c-\Lambda_c \quad\quad\quad\quad\quad.70\cr}\eqn\avwdat$$
 The Regge slope is taken from the
$\rho$--trajectory to be, $\alpha^\prime=.88$\ (GeV)$^{-2}$.  Only those states
which are lowest lying on the respective trajectories are included; the values
for higher lying pairs of states differ only by non--linearities in the
trajectories. To the accuracy given in \avwdat, isospin splittings may be
ignored. Of the pairs listed, five were known to the authors of [\avw].

     The mass squared differences are all (except for the charmed baryons)
surprisingly close to 1/2. This success is particularly puzzling,
 given that {\it apriori}
 the four point amplitudes involved needn't have anything like the simple
 Veneziano form. After all, given the possibilities for conformal field
theories, there are correspondingly rich possibilities for consistent string
amplitudes, all consistent with Regge behavior and duality. The Adler condition
could arise from a simple multiplicative momentum factor and
      it is not difficult to construct amplitudes
with the correct masses which don't satisfy the Adler condition.
As we will now show, however, given the picture for chiral symmetry breaking
we have proposed, the mass
relations of Lovelace and AVW follow even without assuming a Veneziano form for
the four--point amplitudes.

  Consider a three--point function coupling a pion to hadrons
$A$ and $A^*$. This could be included as a factor in an N--point function in
some limit, with either $A$ or $A^*$ appearing as an intermediate state, so we
needn't restrict them to be on the mass shell.  In the limit in which the pion
carries zero momentum the three--point function will vanish because of an
explicit factor of momentum unless the pion has an S--wave coupling to the
other hadrons, so we restrict our attention to this case. The relevent
conformal field theory amplitude, evaluated in the upper--half complex plane
with all vertex operators inserted on the real axis is fixed up to an over all
constant by the conformal symmetry,
$$\eqalign{\langle V_\pi(z,k=0)V_A(x_1,p)V_{A^*}(x_2,-p)\rangle &= \cr
  const.(z-x_1)^{\Delta_{A^*}-\Delta_A-\Delta_\pi}
&(z-x_2)^{\Delta_A-\Delta_{A^*}-\Delta_\pi}(x_1-x_2)^{\Delta_\pi-
\Delta_A-\Delta_{A^*}+
\alpha^\prime p^2}\cr}\eqn\piaa$$

     We have argued that the zero momentum
pion vertex operator should behave like a
holomorphic current in the bulk theory evaluated on the world sheet
boundary.  Under this assumption the amplitude \piaa\ should be well defined if
we analytically continue in $z$ away from the real axis. In particular, \piaa\
should be single valued under the operation of analytically continuing in $z$
around any closed contour. Assuming boundary conditions on the real axis which
preserve isospin symmetry ($J^a(z)=\bar{J}^a(\bar{z})$ for $z$ real), we can
 smoothly continue  the pion operator from the upper half--plane into the
lower half--plane without difficulty. Choosing a contour surrounding both $x_1$
and $x_2$, \piaa\ is single valued under the continuation provided that
$2\Delta_\pi$ is an integer. This is just the restriction that the
world sheet current have
integer or half--integer spin. Imposing single--valuedness for the contour
surrounding $x_1$ alone gives the condition,
$$\Delta_{A^*}-\Delta_A-\Delta_\pi=0\quad\quad ({\rm mod\; 1})\quad
.\eqn\aapi$$
The mass relations found by AVW then immediately follow, with
 $\alpha^\prime(M^2_{A^*}-M^2_A)$ an integer or half--integer depending on the
world sheet spin of the current associated with the pion. The half--integer
spin current (such as occurs in the RNS model) is clearly chosen by the data.
Reconciling this half--integer spin with the fact that the pion is massless,
should ultimately provide a significant clue as to how the intercepts of the
leading trajectories should be shifted in a complete string theory.

We can also  justify, to some extent, the starting point of the Lovelace and
AVW
derivations. In a {\it bulk} conformal field theory
correlation functions involving  holomorphic operators, or more generally any
operators which have simple fusion rules with all other operators, necessarily
have a simple analytic structure. Four--point functions with one or more
such operators can be written as the product of a single
holomorphic function of the
world sheet coordinates times an anti--holomorphic function, with the product
single valued on the complex plane. Open string four--point tree
amplitudes which are constructed from the corresponding boundary operators
are integrals of the holomorphic half of the bulk amplitude; this gives rise to
a finite sum of Veneziano type terms, the form assumed by Lovelace and AVW.
This form is natural,then, for any four--point function involving
pions,\foot{Which may explain why  linear ``zero trajectories''
 are observed in pion--nucleon scattering \refmark{\zerot}} if the other
boundary operators have counterparts in some bulk theory. This is not required
by the picture for chiral symmetry breaking we have been considering.
The boundary operators in a conformal field theory need not have analogs in any
consistent bulk theory. The consistency conditions (duality, modular
invariance, etc.) which correlation functions of bulk or boundary operators
must satisfy, while related, are not identical. If the boundary theory cannot
be extended into a consistent bulk theory (which we believe will be the case at
least for the operators appropriate for describing mesons involving one heavy
 quark) then the four--point functions including pions (with non--zero momenta)
 needn't consist of a finite sum of Veneziano terms. The four pion amplitude
itself is a special case. It should consist of Veneziano terms, and its form is
constrained up to a few free parameters, with the original Lovelace--Shapiro
amplitude the most natural possibility. It is encouraging that the
Lovelace--Shapiro amplitude, with no free parameters, fits the $\pi\pi$
scattering data reasonably well.\refmark{\wip}

    A few further comments on the entries of \avwdat\ and the validity of the
AVW mass relations are in order. First, the derivation we have given is
appropriate for baryons only to the extent that a baryon behaves like a meson
string, for example using either of the approximations for baryons discussed
in section 2. It is entirely reasonable, then, that the observed
agreement of the
baryons with the mass relation is systematically worse than that for the
mesons.
Second, we have been using an approximation in which the pion trajectory
is treated as linear. It is measurably non--linear, and this fact is reflected
in the failure of the AVW mass relation when any of the higher spin states on
the pion trajectory are involved. Finally, there are likely problems with the
higher lying states on the heavy meson trajectories as well, since these are
not expected to be linear, but there is, at present, little data involving
these states to consider.

\chapter{Closed strings, the pomeron, and the additive quark rules for total
hadronic cross sections}

     In any given string model one can analyze the closed string spectrum and
scattering amplitudes as easily as those for open strings.  Were there
compilations of glueball properties and scattering data one could compare
directly with the string results to learn about this sector of the theory,
which is free from complications due to world sheet boundary
effects.  Unfortunately this is not the case.  Only a handful of glueball
candidates exist (none positively identified) and, given that the closed string
Regge slope is half that for open strings, it is too much to expect to identify
several states on a single trajectory.  Indeed there is no evidence that
glueballs are narrow resonances at all.

   The evidence for closed string behavior comes from exchanged t--channel
trajectories in elastic and inclusive processes. At high energies
($s>100$\ GeV$^2$) all total hadronic cross sections become approximately
constant in energy (rising only logarithimically, or as a small power of
$s$). Relating the total cross--section to the forward scattering amplitude via
the optical theorem and assuming Regge behavior, this indicates the exchange of
a trajectory with intercept near one with vacuum quantum numbers, the pomeron.
An approximately linear trajectory with intercept one and slope consistent with
half of the open string value (as predicted for a closed string)
can be isolated in pp elastic scattering at
moderate energies ($P_{{\rm Lab}}\sim 30$\ GeV).\refmark{\data}
At higher energies the measured
slope falls to only 1/3 to 1/4 that for the meson trajectories, consistent with
the flattening of the slope expected from multiple pomeron exchange.

    The fact that the total cross sections do not rise linearly in $s$ is the
most direct evidence that there is no hadronic string trajectory with intercept
two.\foot{In actuality such a linear rise would soon violate the Froissart
bound, so one would expect to observe a near saturation of the Froissart bound
at modest $s$ if there were in fact a trajectory with intercept two.}
In particular any mechanism for eliminating or
decoupling the spin two ``graviton'' is, by itself, insufficient to reconcile
a string theory with hadronic physics if the rest of the trajectory
 remains.\foot{If enough states were absent from the trajectory, however, its
coupling to hadrons at present energies might be sufficiently weak to avoid
conflict with the data.}

      Leaving aside the problem of the pomeron intercept, let us consider the
nature of its couplings to hadrons and determine whether the observed behavior
is consistent with the interpretation of the pomeron as a closed string.  Long
ago it was observed that total hadronic cross sections at high energy
seem to obey an
additive quark rule, i.e., the  total cross--section behaves like a sum of
individual valence quark--quark cross sections.\refmark{\addqk}
To illustrate, for $s$ around 200 GeV$^2$ the measured total cross sections
are,\refmark{\landolt}
$$\sigma_{\pi N},\sigma_{K N},\sigma_{pp},\sigma_{\Sigma N},\sigma_{\Xi
N}= 24,20,39,33,29\ {\rm mb}\eqn\tcross$$
Using the pion--nucleon and kaon--nucleon total cross sections to extract
 cross sections for u or d quarks on u or d quarks and u or d quarks on s
quarks, respectively,  the additivity assumption  leads to the
predictions,
$$\sigma_{pp},\sigma_{\Sigma N},\sigma_{\Xi
N}= 36,32,28\ {\rm mb}\quad ,\eqn\across$$
in reasonable agreement with \tcross.\foot{We have made no attempt here to
remove the small part of the total cross--section at these energies which is
due to non--pomeron exchange although this should be done for a more definitive
test of the additive quark rules.} The agreement for cross sections on deuteron
and helium targets is similar, though some nuclear shadowing is apparent.
 Donnachie and Landshoff, motivated by
this feature of the data, have argued that the pomeron is small by hadronic
standards, and couples locally to the
valence quarks in a hadron rather like a C=+1 photon.\refmark{\donn}

     Within the string picture, the single pomeron exchange contribution to the
forward scattering amplitude of two mesons is given by the open string
non--planar, one--loop (``cylinder'') diagram.  Computing a loop diagram
requires more detailed knowledge of a string theory than a tree amplitude;
problems which we could largely sidestep at tree level,
such as undesirable trajectories or the additional degrees of
freedom required for the cancelation of the conformal anomaly, cannot be
avoided.  Fortunately, in the Regge limit (large $s$, fixed $t$), the
non--planar amplitude
factorizes into a closed string propagator and form factors for the coupling
of the closed string to each of the open strings.\refmark{\lpom,\ales,\cshap}
 The form factors can be
computed at tree--level, so that we can avoid the need for specifying a
complete string model. We will treat the pomeron coupling to mesons first,
before turning to baryons.

     Is the pomeron--meson--meson coupling in the string picture compatible
with the observed additive quark rule? The valence quark
content of a meson is determined by the choice of boundary conditions on the
edge of the string world sheet. The pomeron, on the other hand , is emitted
from the {\it interior} of the meson world sheet, not from the boundaries as
an open string would be, so there seems to be no origin for any simple additive
behavior. By duality we can think of the pomeron as coupling to the meson via
an open string emitted from one of the two meson boundaries which then closes
upon itself (fig.2),
 but this is not the origin of additivity.  Duality equates the three pictures
in fig.2, not the first with the sum of the last two. Since we would expect
the pomeron exchange contribution to the forward scattering amplitude to
involve distance scales for which the string picture is appropriate, this is a
 potentially serious problem for the effective string picture.

To address this issue, let us consider  the pomeron--meson form factor in some
detail. It is a three--point function for two open and one closed string. In
the
underlying conformal field theory this can be written in terms of a correlation
function of two boundary operators and one bulk operator in the upper
half--plane, $\langle \psi_m^{ab}(x_1)\psi_m^{ba}(x_2)\phi_P(z,\bar{z})
\rangle$. The
superscripts $a$ and $b$ label boundary conditions along the real axis; the
boundary operator $\psi_m^{ab}$ connects the two. We will use different
boundary conditions to incorporate different valence quarks.
 The bulk--boundary--boundary three--point function has the same form as the
 holomorphic half of a bulk four--point function in the full complex
plane,\refmark{\cardy}
 $\langle \psi_m^{ab}(x_1)\psi_m^{ba}(x_2)\phi_P(z)\bar{\phi}_P(\bar{z})
\rangle$. Using the conformal invariance of the amplitude we
can fix three real coordinates. The precise prescription requires a little care
because the closed string vertex operator is off mass shell (i.e., not
dimension
(1,1)). In factoring
the complete non--planar four--point function the usual closed string
propagator (which appears in closed string tree amplitudes) is obtained
if we fix one meson vertex operator at the origin, the other at infinity, and
integrate the pomeron vertex over the unit circle,\refmark{\lpom,\ales,\cshap}
 giving the form factor,
$$g_{mmP}^{ab}=\int^\pi_0 d\theta \langle \psi_m^{ab}\mid
\bar{\phi}_P(e^{-i\theta})\phi_P(e^{i\theta})\mid  \psi_m^{ba}
\rangle\quad .\eqn\gpp$$

    $g_{mmP}^{ab}$ is constrained by conformal invariance to have the general
form,
$$\eqalign{g_{mmP}^{ab}&=\int^\pi_0 d\theta \sum_q C^a_{P q}C^{aab}_{qmm}
    G^q(1-e^{-2i\theta})\cr
&=\int^\pi_0 d\theta \sum_r C^b_{P r}C^{bba}_{rmm}
    \tilde{G}^r(1-e^{-2i\theta})\quad .\cr}\eqn\gq$$

     The second equality in \gq\ is a type of duality (fig.2).\refmark{\dcl}
 $C^a_{P q}$ is the operator product coefficient appearing in the short
distance expansion for the pomeron vertex operator near the boundary with
boundary condition $a$,
$$\phi_P(z,\bar{z})\sim(2{\rm Im}z)^{\Delta_q-\Delta_P-\bar{\Delta}_P}
  C^a_{Pq}\psi^{aa}_q({\rm Re}z)+\dots\quad , \eqn\bbope$$
which governs
the amplitude for the closed string $\phi_P$ to transform
into the open string $\psi^{aa}_q$.  $C^{aab}_{qmm}$ is the boundary operator
OPE coefficient giving the open string three--point amplitude for
the emission of $\psi^{aa}_q$ from the boundary with boundary condition $a$ of
 the open string $\psi^{ba}_m$.  The couplings $C^b_{P r}$ and $C^{bba}_{rmm}$
 are the analogous quantities involving the boundary with boundary condition
$b$.                       $G^q$ and $\tilde{G}^r$ are
 analytic functions of the variable $\eta\equiv 1-e^{-2i\theta}$ except at
the possible branch points $\eta=0,\ 1,$ and $\infty$. For positive
$\theta\sim 0$ ($\eta\sim i\epsilon$),
$G^q$ behaves like $\epsilon^{\Delta_q-2\Delta_P}$.
For $\theta\sim \pi$ ($\eta\sim -i\epsilon$),
$\tilde{G}^r$ behaves like $\epsilon^{\Delta_r-2\Delta_P}$. We assume
throughout that the pomeron vertex operator is diagonal
($\Delta_P=\bar{\Delta}_P$).

   To explore the additivity of total cross sections within the string picture
we will impose the  following conditions on $g_{mmP}^{ab}$.
 For each open string
boundary condition (i.e., quark flavor) we assume that the pomeron mixes
directly only with a single, linear, open string trajectory. For u or d quarks
this will be the f, for s quarks the f$^\prime$. For c or b quarks the
trajectories will not be linear, and consequently our results will be suspect
for this case. Further, we assume that the open string couplings are SU(3)
symmetric, so, for example we take $C^{sus}_{KKf^\prime}=C^{usu}_{KKf}=
C^{uuu}_{\pi\pi f}\equiv C_{mmf}$.
This is supported by Regge fits to various measured
two to two scattering processes;\refmark{\data}
the observed SU(3) symmetry breaking is
consistent with the assumption that it arises from shifts of the trajectory
intercepts alone. Finally, we assume that $g^{ab}_{mmP}$ is a smooth function
of $\Delta_q$ and $\Delta_r$ (i.e., of the underlying valence quark masses).

    With these assumptions only a single $q$ and $r$ contribute to \gq, and
$C^a_{P q}G^q=C^b_{P r}\tilde{G}^r\equiv G^{qr}$.
In order for $G^{qr}(1-e^{-2i\theta})$
to have the correct behavior for $\theta$ near 0 and $\pi$ it must be part of a
two-dimensional representation of the monodromy group. That is, the result of
analytically continuing $G^{qr}(\eta)$ about any of its branch points is a
 linear combination of two functions, one behaving like
$\eta^{\Delta_q-2\Delta_P}$  as $\eta\rightarrow 0$ the other
like $\eta^{\Delta_r-2\Delta_P}$. This restricts $G^{qr}(\eta)$
 to be a linear combination of terms involving hypergeometric functions,
of the form,
$$\eta^{\Delta_q-2\Delta_P}(1-\eta)^B[1+\sum^M_{j=1}a_j\eta^j]
       {\rm F}(\alpha,\,\beta;\,\Delta_q-\Delta_r+1;\,\eta)\quad .\eqn\fab$$

This form is further restricted by the following requirements:

1) symmetry under the interchange of $z$ and $\bar{z}$ in the original
amplitude implies $G^{qr}(\eta)=e^{i\pi
(\Delta_q-2\Delta_P)}G^{qr}(\eta/(\eta-1))$;

2) after analytically continuing $G^{qr}(1-e^{-2i\theta})$ from $\theta=0$ to
$\theta=\pi$, only the terms behaving as
$\eta^{\Delta_r-2\Delta_P}$ should appear;

3) the limit in which $q=r$ should be well defined and the result, $G^{qq}$,
 symmetric under the above analytic continuation (i.e., symmetric under the
interchange of $\theta$ and $\pi-\theta$).

The final result (employing transformation and recursion formulas for the
hypergeometric functions as needed and the identity $F(a,1/2+a;1+2a;\eta)=
2^{2a}[1+(1-\eta)^{1/2}]^{-2a}$) is,
$$\eqalign{G^{qr}(\eta)=&C^a_{P q}2^{\Delta_q-\Delta_r}
(-i\eta)^{\Delta_q-2\Delta_P}
(1-\eta)^{\Delta_P-{1\over 4}(\Delta_q+\Delta_r)}\cr
&[1+\sum_{n=1}^N\alpha_n({\rm
cos}2n\theta-1)][1+(1-\eta)^{{1\over 2}}]^{\Delta_r-\Delta_q}\quad .\cr}
\eqn\gqr$$
Here $\alpha_n$ are some linear combinations of the $a_j$ in \fab\ and $N=M/2$
for $M$ even; odd powers are incompatible with condition 3).

   Analytically continuing \gqr\ from $\theta=0$ to $\theta=\pi$ and imposing
the duality symmetry we can relate the two bulk--boundary OPE coefficients
and solve for their  dependence on the boundary condition,
$$4^{\Delta_q}C^a_{P q}=4^{\Delta_r}C^a_{P r}\equiv
4^{-\alpha^\prime t}\kappa\quad .\eqn\ck$$
$\kappa$ is the bulk--boundary OPE coefficient in the simple bosonic string,
and $t$ is minus the pomeron momentum squared.
Together with \gqr\ this gives the general form for the pomeron--meson--meson
form factor \gpp\ consistent with world sheet conformal invariance, duality,
and the assumption of pomeron mixing through a single meson trajectory for each
boundary condition,
$$g_{mmP}^{ab}=\int^\pi_0 d\theta\ 2^{-\Delta_q-2\Delta_P-2\alpha^\prime t}
\kappa C_{mmf}
           [1+\sum_{n=1}^N\alpha_n({\rm cos}2n\theta-1)]
 ({\rm sin}\theta)^{\Delta_q-2\Delta_P}
            ({\rm cos}\theta/2)^{\Delta_r-\Delta_q}\quad .\eqn\gthet$$
 With a change of variables, $x={\rm sin}^2\theta/2$, we can bring this
expression into a more symmetric form and integrate it,
$$\eqalign{g_{mmP}^{ab}&=\!\int^1_0 \!dx\kappa C_{mmf}
          2^{-4\Delta_P-2\alpha^\prime t}
  x^{{1\over 2}(\Delta_q-2\Delta_P-1)} (1-x)^{{1\over 2}(\Delta_r-2\Delta_P-1)}
         \bigl[1\!+\!\!\sum_{n=1}^N a_n((1-2x)^{2n}\!\!-\!1)\bigr] \cr
   &=\kappa C_{mmf} 4^{\alpha_P(t)-2-\alpha^\prime t}
{\rm B}\bigl({\scriptstyle {1\over 2}}(\alpha_P(t)-\alpha_q(t)),\,
 {\scriptstyle {1\over 2}}(\alpha_P(t)-\alpha_r(t))\bigr)\cdot
\cr &\phantom{=\kappa C_{m}}
\bigl[1+\sum_{n=1}^N a_n\bigl({\rm F}\bigl(-2n,\,
 {\scriptstyle {1\over 2}}(\alpha_P(t)-\alpha_r(t));\, {\scriptstyle {1\over
2}}
(2\alpha_P(t)-\alpha_q(t)-\alpha_r(t));\,2\bigr)-1\bigr)\bigr]\, .
\cr}\eqn\gx$$
 B is the Euler beta function,
the coefficients $a_n$ are linear combinations
of the $\alpha_n$ in \gthet, and in the last expression we have replaced the
conformal dimensions by the related Regge trajectory functions,
 $$\eqalign{1-\Delta_q&=\alpha_q(t)=\alpha^\prime t+\alpha_q\cr
            1-\Delta_r&=\alpha_r(t)=\alpha^\prime t+\alpha_r\cr
           2-2\Delta_P&=\alpha_P(t)={\alpha^\prime\over 2} t
+\alpha_P\quad .\cr}\eqn\da$$
The hypergeometric function represents a finite sum of ratios of polynomials of
the trajectory functions since the first argument is a negative integer.

    In the limit $q=r$ with $N=0$, \gx\ reproduces the pomeron form factor
first found by Lovelace for the Veneziano model.\refmark{\lpom}
 The same expression is valid for the Neveu--Schwarz model as
well.\refmark{\cshap} Equation \gx\ with $N=0$ is
the direct generalization of this result, consistent with conformal
invariance and duality, for pomerons coupling to meson strings
with different ``quarks'' on either end. As one would expect, a change in one
of the string boundaries has two effects: the conformal dimension $\Delta_q$
of the
boundary state appearing in the limit that the pomeron vertex operator
approaches the boundary is altered, as is the relevent coupling, $C^a_{Pq}$.
 Because of the duality constraint the two
effects are completely correlated (c.f.,\ck).

   We can finally test whether the additive quark rules for total cross
sections
arise within the string picture. Approximate additivity would mean that for
vanishing pomeron momentum,
$$g_{mmP}^{ab}\approx {1\over 2}(g_{mmP}^{aa}+g_{mmP}^{bb})\quad .\eqn\addit$$
We first consider the minimal case (\gx\ with $N=0$) for which
\addit\ becomes,
$${\rm B}({\scriptstyle {1\over 2}}(\alpha_P-\alpha_q),\,
 {\scriptstyle {1\over 2}}(\alpha_P-\alpha_r))
\approx {1\over 2}[{\rm B}({\scriptstyle {1\over 2}}(\alpha_P-\alpha_q),\,
 {\scriptstyle {1\over 2}}(\alpha_P-\alpha_q))
 +{\rm B}({\scriptstyle {1\over 2}}(\alpha_P-\alpha_r),\,
 {\scriptstyle {1\over 2}}(\alpha_P-\alpha_r))]\eqn\addb$$
The case of physical interest (to explain \across) is $q$ and $r$ taken as the
f and f$^\prime$ mesons with intercepts near .5 and .1 respectively, together
with a pomeron intercept of 1. For these values \addb\ holds to better than
${\scriptstyle {1\over 2}}$\%. In fact for any values of these intercepts
within the range from
$0< \alpha_q,\alpha_r <1$, \addb\ holds to better than 1\%. For a range between
$-.9$ and 1 the agreement is
still better than 5\%, and better than 10\% even for very large negative
intercepts (where our approximation of linear trajectories is in fact no longer
trustworthy) provided $|\alpha_r-\alpha_q|<1$. As an aside, for f and
f$^\prime$
intercepts of .5 and .1 respectively, the minimal form factor predicts
$\sigma_{KN}/\sigma_{\pi N}=.74$, in rough agreement with \tcross.

    Numerically checking \addit\ for subsequent terms in \gx\
($n$=1,2,3\dots), one finds that as $n$ increases
 the additivity approximation for the individual terms improves and the
contribution to the form factor slowly drops. Thus if the coefficients $a_n$
are independent of $\Delta_q$ and $\Delta_r$ (or even slowly varying
 functions), and very large cancelations among the coefficients are
avoided, then
 the general form factor $g^{ab}_{mmP}$ is remarkably additive,
in the sense of \addit, over a broad range of the parameter space which
includes the physically relevant values for the intercepts.

    Why does \addit\ hold so well for \gx? It is not simply that the dominant
contributions to the form factor come from pomeron emission near the world
sheet boundaries. This is true for $\alpha_q$ near 1 (where $g_{mmp}^{ab}$
becomes infinite), but not for smaller $\alpha_q$ where approximate additivity
still holds. The success of \addit\ follows also from the structure of the
underlying bulk--boundary--boundary three--point amplitude. This is a smooth
analytic function of
$\theta$ between 0 and $\pi$ except at the end points, where the holomorphic
half of the pomeron vertex operator collides with its anti--holomorphic image
at the world sheet boundary. Thus the endpoints play a special role; for
$\theta$ away from 0 and $\pi$ the amplitude is essentially determined
by the behavior at the nearest branch point,
with a smooth interpolation in the middle around $\theta=\pi/2$.
As we consider terms in \gx\ with increasing $n$, the contribution of pomeron
emission from the middle of the world sheet becomes less and less dependent on
the behavior at the boundaries because
of the factor cos$^{2n}\theta$, further improving the additivity
approximation for these terms.

    We turn now to the form factor for pomeron emission from baryons. Ideally
we would compute the conformal field theory amplitude for the pomeron vertex
operator to be emitted from the interior of a world sheet consisting of three
surfaces glued together along a curve (fig.1). We would then have to integrate
 this amplitude over some moduli space of   joining curves as well as
over the position of the pomeron operator. We would expect additivity to follow
for the same heuristic reason given above for the meson case: the behavior of
the amplitude should be governed by the singularities arising when the pomeron
vertex operator approaches the world sheet boundary (which consists here
of three
lines joined at a point). Elsewhere the amplitude should  be a smooth function
 of the position of the pomeron vertex operator, even on the joining
curve. Unfortunately we lack the necessary
technology for such a complete computation. We will consider the form factor
 only within the two approximations for baryons discussed in section 2.

    Consider, then, $g^{abc}_{BBP}$, the form factor for pomeron emission
from a baryon world sheet with boundary conditions $a,b,c$ on the three
distinct edges.  For that part of the amplitude in which
the pomeron vertex operator is
located on the subsheets of the world sheet bounded by the $a$ and $b$
boundaries, we can integrate over the possibilities for the third subsheet
with boundary $c$ in order to reduce the problem to the same form as the meson
case. Because we have integrated out this third subsheet, the conformal field
theory on the new (meson--like) world--surface could be different from that in
the pure meson case; however, the leading behavior of the amplitude for the
pomeron near the boundary will be the same since $\Delta_P,\ \Delta_q$, and
$C^a_{P q}$ remain unchanged. Thus this contribution to the form factor is of
the same form as in the meson case, $g^{ab}_{mmP}$.

   To include the contribution to the form factor coming from pomeron emission
from the subsheet bounded by $c$, we can integrate out either the $a $ or $b$
subsheets.  Clearly the complete form factor for baryons within this
approximation is,
$$g^{abc}_{BBP}={1\over 2}(g^{ab}_{mmP}+g^{bc}_{mmP}+g^{ac}_{mmP})\quad
.\eqn\gbbp$$
The additivity of this form factor follows from the additivity of the meson
form factor provided that the coefficients $a_n$ are constant or change only
slightly as we move from the meson case to the baryon case.

    In the second approximation to baryons within the string picture, we
consider an open string world sheet with a ``quark'' on one boundary and a
``diquark'' on the other, represented by boundary conditions $a$ and $bb$
respectively.  For simplicity we will take the two quarks within the diquark to
be degenerate, which we can always do if we restrict our attention to u, d, and
s quarks and assume exact isospin symmetry. In order to incorporate three
distinct quark masses we would have to modify the form
factor to include functions transforming as a three dimensional representation
of the monodromy group.

  The computation of the form factor follows that for the meson case except for
two features.  In the short distance expansion for the pomeron vertex operator
near the $bb$ diquark boundary the coefficient of the boundary operator
$\psi^{bb}_r$ should be twice what it was for the single $b$ boundary,
$C^{(bb)}_{Pr}=2C^b_{Pr}$. Second, the form factor is not symmetric under the
interchange $\theta\leftrightarrow\pi -\theta$ even in the limit
$\Delta_q=\Delta_r$, since the two boundaries are inherently different. This
means we must allow for cos$j\theta$ terms in \gqr\ with $j$ odd as well as
even, and in fact such terms are required in order to achieve
$C^{(bb)}_{Pr}=2C^b_{Pr}$. The final structure of the the form factor for
baryons within this approximation is,
$$\eqalign{g_{BBP}^{abb}&=\int^1_0 dx\kappa C_{mmf}
   2^{-4\Delta_P-2\alpha^\prime t}
  x^{{1\over 2}(\Delta_q-2\Delta_P-1)} (1-x)^{{1\over 2}(\Delta_r-2\Delta_P-1)}
\cr&\phantom{=\int^1_0 dx\kappa C_{mmf}}
\cdot\Bigl[1+x+\sum_{n=1}^N[b_n+c_n(1-2x)]((1-2x)^{2n}-1)\Bigr] \cr
   =\kappa &C_{mmf}4^{\alpha_P(t)-2-\alpha^\prime t}
{\rm B}({\scriptstyle {1\over 2}}(\alpha_P(t)-\alpha_q(t)),\,
 {\scriptstyle {1\over 2}}(\alpha_P(t)-\alpha_r(t)))
\Bigl[1+{\alpha_P(t)-\alpha_q(t)\over
2\alpha_P(t)-\alpha_q(t)-\alpha_r(t)}
\cr &+\sum_{n=1}^N a_n({\rm F}(-2n,\,
 {\scriptstyle {1\over 2}}(\alpha_P(t)-\alpha_r(t));\,{\scriptstyle {1\over 2}}
(2\alpha_P(t)-\alpha_q(t)-\alpha_r(t));\,2))-1)\cr
&\phantom{\sum_{n=1}^N}+b_n({\rm F}(-2n-1,\,
{\scriptstyle {1\over 2}}(\alpha_P(t)-\alpha_r(t));\,{\scriptstyle {1\over 2}}
(2\alpha_P(t)-\alpha_q(t)-\alpha_r(t));\,2)\cr
&\phantom{\sum_{n=1}^N}-1+2{\alpha_P(t)-\alpha_q(t)\over
2\alpha_P(t)-\alpha_q(t)-\alpha_r(t)})\Bigr]\quad . \cr}\eqn\gbx$$

   For the minimal form factor ($N=0$) at zero pomeron momentum this reduces
to,
$$g^{abb}_{BBP}=g^{ab}_{mmP}[1+{\alpha_P-\alpha_q\over
2\alpha_P-\alpha_q-\alpha_r}]\quad .\eqn\gbmin$$
 For $q=r$ this satisfies the additivity condition exactly (as does \gbbp),
$g^{aaa}_{BBP}={\scriptstyle {3\over 2}}g^{aa}_{mmP}$.  For $\alpha_q$ and
$\alpha_r$ within the physically interesting range $-.5<\alpha_q,\ \alpha_r<1$
the additivity property and the agreement between the two approximations
\gbmin\ and \gbbp\ for the baryon form factor, hold to 5\% or better.
There is no unambiguous choice for the higher order coefficients $b_n$ and
$c_n$ in terms of the coefficients $a_n$ appearing in the meson form factor,
but it should be such that the behavior of the integrand near $x=0$ and $x=1$
approximates the behavior of the integrand in the meson form factor near $x=0$
and twice the integrand near $x=1$, respectively. Approximate additivity
again results.

\chapter{Conclusions}

    In this work we have advocated the use of effective string amplitudes for
hadron phenomenology. This subject is an old one --- the earliest days of Dual
models included some surprisingly succesful attempts at phenomenolgy using the
Veneziano and Koba--Nielsen amplitudes --- but today a more complete and
systematic approach is possible using the technology of conformal field
theories.\refmark{\bpz} The general structure and properties of amplitudes
consistent with world sheet conformal invariance, duality and Regge behavior
are well understood; examples are no longer restricted to the rather special
case of Veneziano type models. The current challenge is finding the best means
for incorporating various properties of hadrons within this framework. We
believe in particular that the (under utilized) technology of conformal field
theory amplitudes on surfaces with boundaries together with appropriately
chosen boundary conditions,\refmark{\cardy, \dcl} will make it possible to
incorporate the properties of the valence quark content of hadrons.

   The natural subjects for this approach are soft hadronic processes for which
perturbative QCD is inapplicable and lattice gauge theory is often impractical.
This should include soft hadronic contributions to weak processes (e.g., heavy
meson decay constants, or the $\Delta$I=1/2 rule) as well as physics in the
Regge regime, diffractive (pomeron) processes and hadron spectroscopy. The
imposition of string behavior (linear trajectories, world sheet conformal
invariance, duality) is highly constraining and reduces the number of free
parameters available in traditional Regge phenomenology dramatically. Perhaps
more important than these applications are the possibilities for improved
qualitative understanding of inherently non--perturbative gauge theory
phenomena (such as chiral symmetry breaking or the nature of the pomeron) which
may have applications ranging from lattice gauge theory to technicolor.

   We have considered two applications of these ideas to begin to
illustrate the possibilities. In the first we proposed a natural mechanism for
chiral symmetry breaking within the string picture based on a generic feature
of conformal field theories: that some of the symmetries present in the bulk
theory are necessarily broken by the choice of boundary conditions when
boundaries are included. In this picture, chiral symmetry breaking necessarily
follows from confinement and the existence of mesons. One consequence of this
picture is that the vertex
operators for pion emission are particularly special, behaving (at zero
momentum) like world sheet current operators evaluated on the string boundary.
We used this fact, in turn, to rederive, under more general conditions, the
hadronic mass relations first found by Lovelace\refmark{\lovelace} and
Ademollo,
Veneziano and Weinberg.\refmark{\avw}. Even though the present derivation still
includes a number of assumptions and the original suggestion is more than 20
years old, this remains  a remarkable result:
string behavior together with chiral symmetry essentially predicts the correct
$\rho$ mass, as well as other mass relations.

   In the second example we derived the general structure of the form factor
for emission of a pomeron (interpreted as a closed string) from a meson or
baryon, under modest assumptions. The result generalizes that found for the
bosonic and RNS strings.\refmark{\lpom,\ales,\cshap} Remarkably, within a
broad parameter range including the physically most relevant values for the
trajectory intercepts, these form factors display an additive behavior, as
required to reproduce the additive quark rules for total hadronic cross
sections which are observed experimentally.
To the extent that QCD hadrons behave like flux tubes, this represents, to our
knowledge, the first explanation in the completely nonperturbative regime of
QCD for the success of the additive quark rules.

  This result is not necessarily in conflict with Donnachie and Landshoff's
phenomenological explanation of the additive quark rule in which the pomeron
couples locally to the valence quarks.
 In the string picture pomeron emission from the interior of
the string world sheet away from the boundaries represents a significant
contribution to the form factor. Because of conformal invariance, however, it
 is not clear which
configurations of the string couplings in {\it space--time} provide the
dominant contributions. While our results provide an alternative origin for the
additive quark rule, it is possible that even in the string picture the pomeron
effectively couples locally to the string boundaries as measured in
space--time.

    The emergence of the additive quark rule from the string picture provides
some insight into why the naive quark model is often successful even for
processes in which sea quarks and gluons are expected to be important.  In the
string picture the boundary conditions on the edge of the world sheet, together
with the behavior of the bulk operators as they approach the boundary, to a
large extent govern the behavior of simple string amplitudes, even those, such
as pomeron emission, where the interior of the world sheet (presumably built up
from the contributions of gluons and sea quarks) is involved in a fundamental
way.

    Admittedly, settling for effective string amplitudes instead of complete
string theories is less than completely satisfactory.  Indeed one of the chief
motivations for this approach is to gain insight into what sort of string
behavior is most relevant for hadron physics as a step towards constructing a
complete theory. At this stage we are restricted to tree amplitudes and
therefore processes for which the narrow resonance approximation is
appropriate.
Further, the approximation of string--like behavior for hadrons as embodied
in conformal field theory amplitudes, is completely uncontrolled. Judging from
the measured linearity of hadron trajectories and the degree of success of
earlier Regge and Dual model phenomenology, we expect results valid to 5--20\%
accuracy for hadronic processes dominated by string--like behavior, but there
is no understanding of how to compute corrections to these results or determine
precisely when the approximations break down. This will require new technology
for summing over world sheets in the absence of conformal invariance.
 The present situation is in many ways analogous to the early days of current
algebra prior to the systematic exploitation of effective chiral Lagrangians,
only in this case we do not yet know how to handle the analog of the field
theory with explicitly broken chiral symmetry.

   Ultimately we would like to derive an effective string theory directly from
QCD. This is  particularly difficult because it involves comparing a second
quantized theory (QCD) in which we can't compute hadron amplitudes, with a
first quantized theory (string theory) in which basically all we can do is
 compute S--matrix elements perturbatively.  A different formulation of one or
both theories is required to even compare the two on the same footing.
 We have avoided addressing the question of whether
 QCD is in some limit {\it exactly} equivalent to some string theory,
and if so how  hard parton--like behavior  arises in the string picture.
We feel it is likely that the effective theory which best
describes the observed string--like behavior of hadrons
is not the the same as the exact
string equivalent to some limit of QCD (assuming that one exists).
Nonetheless, the former would provide great insight into the latter as well as
providing a valuable tool for phenomenology.

   We would like to thank colleagues at SLAC, the Institute for Theoretical
Physics and the University of California at Santa Barbara,
in particular M.~Alford, J.~Cardy, M.~Einhorn, M.~Peskin and L.~Susskind,
for useful discussions. This research was supported by the National Science
Foundation under Grant No.  PHY 89-04035.

\endpage
\refout
\endpage
\FIG\one{A baryon world sheet.}
\FIG\two{The emission of a closed string from an open string can, by duality,
be computed as a sum over open string states (emitted from {\it one} of the two
boundaries) which then transform into the closed string.}
\figout
\endpage
\end